\documentclass[authoryear,5p,twocolumn]{elsarticle}
\usepackage{epsfig}
\usepackage{color}
\usepackage{amssymb}
\usepackage{amsmath}
\usepackage{amsthm}
\usepackage{multirow}
\usepackage{twoopt}
\usepackage{graphicx}
\definecolor{slateblue}{rgb}{0.1,0.22,0.58}
\usepackage[pdftex,breaklinks=true,colorlinks=true,linkcolor=slateblue,citecolor=slateblue,urlcolor=slateblue,pdfauthor={Tomassetti},pdftitle={SPHe}]{hyperref}
\usepackage{xspace}

\newcommand{\ApJ}{Astrophys. J.}
\newcommand{\AeA}{Astron. \& Astrophys.} 
\newcommand{\PRL}{Phys. Rev. Lett.}
\newcommand{\PRD}{Phys. Rev. D}
\newcommand{\PRC}{Phys. Rev. C}
\newcommand{\ASR}{Adv. Space Res.}

\newcommand{\MNRAS}{MNRAS}

\newcommand{\AMS}{\textsf{AMS}\xspace}
\newcommand{\etal}{et al.}
\newcommand{\eg}{\textit{e.g.}} 
\newcommand{\ie}{\textit{i.e.}} 

\newcommand{\p}{\textsf{p}\xspace}
\newcommand{\Hyd}{\textsf{H}\xspace}
\newcommand{\He}{\textsf{He}\xspace}
\newcommand{\pHe}{\textsf{p/He}\xspace}
\newcommand{\Li}{\textsf{Li}\xspace}
\newcommand{\Be}{\textsf{Be}\xspace}
\newcommand{\B}{\textsf{B}\xspace}
\newcommand{\C}{\textsf{C}\xspace}

\newcommand{\Fe}{\textsf{Fe}\xspace}

\newcommand{\BC}{\textsf{B}/\textsf{C}\xspace}  

\def\pbar{$\overline{p}$\xspace}

\def\ep{$e^{+}$\xspace}
\def\Journal#1#2#3#4{{#4}, {#1}, {#2}, #3} 

\newcommand*{\doi}[1]{\href{http://dx.doi.org/#1}{doi: #1}}

\makeatletter
\journal{Advances in Space Research} 

\begin{document}
\begin{frontmatter}
\title{Testing universality of cosmic-ray acceleration \\with proton/helium data from AMS and Voyager-1} 
\author[pg,lpsc]{Nicola Tomassetti\corref{cor}}
\ead{nicola.tomassetti@cern.ch}
\address[pg]{Universit\`a degli Studi di Perugia \& INFN-Perugia, I-06100 Perugia, Italy} 
\address[lpsc]{LPSC, Universit\'e Grenoble-Alpes \& CNRS/IN2P3, F-38026 Grenoble, France}

\begin{abstract}
The Alpha Magnetic Spectrometer (\AMS) experiment onboard the International Space Station (ISS) has recently
measured the proton and helium spectra in cosmic rays (CRs) in the GeV-TeV energy region. 
The spectra of proton and helium are found to progressively harden at rigidity $R=pc/Ze \gtrsim$\,200 GV,
while the proton-to-helium ratio as function of rigidity is found to fall off steadily as \pHe$\propto R^{-0.08}$.
The decrease of the \pHe{} ratio is often interpreted in terms of particle-dependent acceleration, 
which is in contrast with the universal nature of diffusive-shock-acceleration mechanisms.
A different explanation is that the \p-\He anomaly originates from a flux transition between two components:
a sub-TeV flux component ($L$) provided by hydrogen-rich supernova remnants with soft acceleration spectra, and a multi-TeV
component ($G$) injected by younger sources with amplified magnetic fields and hard spectra. In this scenario the universality
of particle acceleration is not violated because both source components provide composition-blind injection spectra.
The present work is aimed at testing the universality of CR acceleration using the low-energy part of the CR flux,
which is expected to be dominated by the $L$-type component. 
However, at kinetic energy of $\sim$\,0.5-10\,GeV, the CR fluxes are significantly affected by energy losses and solar modulation,
hence a proper modeling of Galactic and heliospheric propagation is required. 
To set the key properties of the $L$-source component, I have used the Voyager-1 data collected in the interstellar space.
To compare my calculations with the \AMS data, I have performed a determination of the force-field modulation parameter using neutron monitor measurements.
I will show that the recent \p-\He data reported by \AMS and Voyager-1 are in good agreement with the predictions of such a scenario,
supporting the hypothesis that CRs are released in the Galaxy by universal, composition-blind accelerators. 
At energies below $\sim$\,0.5\,GeV/n, however, the model is found to underpredict the data collected by PAMELA from 2006 to 2010.
This discrepancy is found to increase with increasing solar activity, reflecting an expected breakdown of the force-field approximation.
\end{abstract}
\end{frontmatter}
\parindent=0.5 cm

\section{Introduction}      
\label{Sec::Introduction}   
%
Proton and Helium nuclei are the most abundant components of Galactic cosmic rays (CRs).
They are mostly accelerated in supernova remnants (SNRs) up to $E\gtrsim$\,1000\,TeV energies before being released into the Galaxy. 
During propagation through the turbulent interstellar medium (ISM), their energy spectra and composition are significantly modified
by diffusion, energy changes, and interactions with the gas nuclei of the ISM.
High-energy collisions of protons and \He nuclei give rise to secondary particles, such as $^{2}$\Hyd, $^{3}$\He, \pbar or \ep,
which bring valuable information on CR propagation.
Besides their interstellar propagation, CRs reaching the Earth are also affected by the solar wind in its embedded magnetic field, 
which \emph{modulates} the shape of their energy spectra below $\sim$\,10\,GeV/nucleon energy.
Hence the interpretation of the data requires a detailed modeling of CR acceleration and propagation processes,
including the time-dependent solar modulation effect \citep{Blasi2013,Grenier2015,Potgieter2013}.
In recent years, new-generation experiments of CR detection have reached an unmatched level of precision
that permits the investigation and eventual resolution of longstanding questions in CR physics \citep{Maestro2015,Serpico2015}.
For instance, the recent measurements of high-energy proton and \He operated by PAMELA and \AMS have revealed the 
appearance of unexpected features in their spectrum \citep{Adriani2011,Aguilar2015Proton,Aguilar2015Helium}.
Here I refer to the so-called proton-to-helium (\pHe) anomaly, \ie, the unexplained spectral difference between 
protons and helium, and to the observation of a common spectral hardening occurring in both particle fluxes at high energy.
More precisely, the \AMS Collaboration reported that the proton and \He fluxes progressively hardens at 
rigidity ($R=pc/Ze$) larger than $R\sim$\,200 GV. The rigidity dependence of the \He flux is similar 
to that of the proton flux, but the helium flux is systematically harder than the proton flux.
Remarkably, the spectral index of the \pHe ratio above 45 GV is well described by a single power-law in rigidity,
$\sim R^{\Delta}$, with $\Delta\sim$\,0.077$\pm$0.007. 
These results pose a serious challenge to standard models of diffusive-shock-acceleration, as they were believed
to be $Z$-independent rigidity mechanisms giving universal power-law spectra \citep{Schwarzchild2011,Serpico2015}
Different models for the origin of the features were proposed in terms of acceleration or diffusion mechanisms.
For instance, the \p-\He spectral difference can be ascribed to CR sources due to non-uniformity of the matter in the acceleration
environment \citep{ErlykinWolfendale2015}, in possible combination with a time-dependent acceleration  \citep{OhiraYoka2011}. 
In \citet{Malkov2012}, it is argued that harder \He spectra may arise from a preferential \textsf{He}$^{2+}$ injection 
occurring in strong shocks. Finally, \citet{Fisk2012} proposed an elemental-dependent acceleration process
occurring in the interstellar turbulence through a series of adiabatic compressions and expansions. 
These mechanisms for the \pHe anomaly, all based on intrinsic CR acceleration, have two main features.
First, the \pHe ratio is expected to decrease steadily at all energies, from sub-GeV to multi-TeV and beyond.
Second, these mechanisms do not automatically explain the spectral hardening in the single CR fluxes, which may be ascribed to 
acceleration or propagation effects \citep{Tomassetti2012Hardening,Tomassetti2015TwoHalo,Aloisio2015}, or to superposition 
of local and distant sources \citep{TomassettiDonato2015,ThoudamHorandel2013,Bernard2013,ErlykinWolfendale2012}.

In \citet{Tomassetti2015PHeRatio}, the decrease of the \pHe ratio was interpreted as a flux transition between 
two source components characterized by different spectra and composition. In this scenario
(hereafter TCM, \emph{two-component model}) the bulk of the $\sim$\,GeV--TeV flux
is ascribed to hydrogen-rich sources with soft acceleration spectra,  while the TeV-PeV flux 
is provided by younger and brighter SNRs with amplified magnetic fields and harder spectra. 
Remarkably, the ``universality'' of the acceleration spectra is not violated in this model,
since each class of source provides elemental-independent injection spectra. As shown, this simple idea can explain both the 
\pHe anomaly and the spectral hardening in proton and \He fluxes. According to this model, the anomalous \pHe{} behavior must
asymptotically disappear at high ($R\gtrsim$\,1000\,GV) and low ($R\lesssim$\,10\,GV) rigidity, \ie, where 
the flux reflects the properties of only one class of contributing sources.
In particular in the high-rigidity region where energy losses are negligible, the ratio must flatten. 
As discussed in \citet{Tomassetti2015PHeRatio}, such a flattening is hinted at by existing multi-TeV data 
and will be resolutely tested by forthcoming CR detection experiments.
In this paper I turn the attention on the low-energy part of the spectrum, down to $E\sim$\,100\,MeV,
where the flux is provided by the hydrogen-rich source component and the injection \pHe{} ratio is expected to flatten.
To provide flux calculations at these energies, it is important to account for elemental-dependent effects such as
energy losses and solar modulation \citep{Potgieter2013,Wiedenbeck2013}. 
In CR astrophysics, solar modulation parameters are often determined by fitting the same sets of data 
that are used to infer the parameters of CR propagation in the Galaxy.
This approach introduces strong degeneracies between modulation and injection/transport parameters.
Models of solar modulation have also been inadequately described because of limited knowledge 
of what exactly the local interstellar spectra (IS) of CRs below a few GeV are outside the heliosphere 
and inadequate continuous observations over an extended energy range of oppositely charge CRs such 
as electrons and positrons, protons and antiprotons.
In this respect, the accomplishments of strategic space missions in the course of the last years are enabling us to make significant progress. 
With the entrance of Voyager-1 in the interstellar space, beyond the influence of the solar wind,
the direct comparison of computed Galactic spectra with experimental data has become possible \citep{Stone2013,Potgieter2014}.
With the PAMELA experiment in space since 2006 and the \AMS installation on the ISS in 2011, 
long-term monitors of particle/antiparticle fluxes have become available \citep{Adriani2016,Consolandi2015}. 
These observations add to a large wealth of data on electrons and nuclei collected over the last decades
by space missions such as CRIS/ACE, IMP-7/8, \emph{Ulysses}, or EPHIN/SOHO \citep{GarciaMunoz1997,Heber2009,Wiedenbeck2009,Kuhl2016},
as well as from the counting rates provided by the neutron monitor (NM) worldwide network \citep{Steigies2015}.

In this paper, the TCM will be tested with the new \AMS measurements on proton and \He in combination with the recent data from Voyager-1.
To characterize the strength of CR modulation over the \AMS observation period (May 2011 - November 2013), I will make use of complementary
sets of data provided by NM stations. This approach mitigates the problem of degeneracy between modulation and 
injection/propagation effects, where the latter constitute my main subject of investigation.
Besides, I will address the problem of modeling solar modulation to describe CR data collected over large 
observation periods (such as the \AMS data that are collected over 2.5 years) during which the solar activity evolves appreciably.
This paper is organized as follows.
In Sect.\,\ref{Sec::CRPropagation} I describe the calculations for Galactic CR propagation in the ISM.
In Sect.\,\ref{Sec::SolarModulation} I describe the modeling of solar modulation and its application to the \AMS data.
In Sect.\,\ref{Sec::Results} I discuss the results and the main focus points of this work.
Conclusions are drawn in Sect.\,\ref{Sec::Conclusions}.

\section{CR acceleration and propagation calculations}  
\label{Sec::CRPropagation}                              

The transport of CRs in the ISM is dominated by particle diffusion in magnetic turbulence and interactions with the matter, 
that is described by a diffusion-transport equation including injection functions,diffusion coefficient, energy losses, 
nuclear interactions and decays \citep{Grenier2015}. The diffusion-transport equation for a $j$-type CR particles is given by: 
\begin{equation}\label{Eq::PropagationEquation}
  \frac{\partial \mathcal{N}_{j}}{\partial t} = q^{tot}_{j}+\vec{\nabla}\cdot\left(D\vec{\nabla}\mathcal{N}_j\right) - {\mathcal{N}_j}{\Gamma^{tot}_{j}} \nonumber - \frac{\partial}{\partial p}\left(\dot{p}_{j} \mathcal{N}_{j}\right) 
\end{equation}
where $\mathcal{N}_{j}=dN_{j}/dVdp$ is the phase space density of the $j$-th CR species.
This equation reflects the most essential features of CR transport. It can be numerically solved for a 
cylindrical diffusive region once interstellar gas and source distributions are specified.
Models of CR propagation in the Galaxy employ analytical or semi-analytical calculations 
\citep{Maurin2001,ThoudamHorandel2013,TomassettiDonato2012},
or fully numerical solvers \citep{Grenier2015,StrongMoskalenko1998}.
The present work relies on numerical calculations under a plain diffusion model of CR propagation
implemented under \texttt{GALPROP} \citep{StrongMoskalenko1998,Vladimirov2011}.
In Eq.\,\ref{Eq::PropagationEquation}, the source term is $q^{\rm tot}_{j}=q^{\rm pri}_{j} + q^{\rm sec}_{j}$, including the 
primary acceleration spectrum (from SNRs) and the term arising from the secondary production in the ISM or decays. 
The primary spectrum is $q_{j}^{\rm pri} = q_{j}^{0}\left(R/R_{0}\right)^{-\nu}$, is normalized to 
the abundances, $q_{j}^{0}$, at the reference rigidity $R_{0}\equiv$\,2\,GV. 
In this work I have implemented a TCM scenario with two diffuse components arising from two classes of primary sources 
with spectral indices $\nu=2.6$ for the low-energy $L$-component and $\nu=2.1$ for the high-energy $G$-component.
For each class of source, the injection spectral indices are imposed to be the same for all the primary elements,
reflecting the \emph{universality of the acceleration spectra}. 
Under \texttt{GALPROP}, the primary composition factors are tuned to the available CR data. 
The source abundances, however, are in general not universal. As shown in \citet{Tomassetti2015PHeRatio}, 
a combination of slightly different composition factors $q_{j}^{0}$ for the two accelerators
may explain the GeV-TeV decrease of the \pHe ratio under a scenario with universal acceleration spectra.
From the data, one has about $q_{\rm He}^{0}/q_{\rm H}^{0}\approx$\,10/90 for the $L$-component,
and  $q_{\rm He}^{0}/q_{\rm H}^{0}\approx$\,18/82 for the $G$-component. The secondary production term is 
$q_{j}^{sec} = \sum_{k} \mathcal{N}_{k} \Gamma_{k\rightarrow j}$, describing the products of decay and spallation 
of heavier CR progenitors with density $\mathcal{N}_{k}$. 
The rate of secondary production $k\rightarrow j$ from CR collisions with the interstellar gas is:
\begin{equation}\label{Eq::ProductionRate}
  \Gamma_{k\rightarrow j} =  \beta_{k}c \sum_{\rm ism} \int_{0}^{\infty} n_{\rm ism} \sigma_{k\rightarrow j}^{\rm ism}(E,E') dE'\,, 
\end{equation} 
where $n_{\rm ism}$ are the number densities of the ISM nuclei, $n_{H}\approx 0.9\,cm^{-3}$ and $n_{He}\approx 0.1\,cm^{-3}$,  
and $\sigma_{k \rightarrow j}^{\rm ism}$ are the fragmentation cross-sections for the production of a $j$-type species at energy 
$E$ from a $k$-type progenitor of energy $E'$ in \Hyd or \He targets. 
$\Gamma^{\rm tot}_{j}=\beta_{j}c \left( n_{\rm H}\sigma_{j,{\rm H}}^{\rm tot} + n_{\rm He}\sigma_{j,{\rm He}}^{\rm tot} \right) + \frac{1}{\gamma_{j}\tau_{j}}$ 
is the total destruction rate for inelastic collisions (cross section $\sigma^{\rm tot}$) and/or 
decay for unstable particles with lifetime $\tau_{j}$.
To handle the nuclear reaction network, Eq.\,\ref{Eq::PropagationEquation} is solved starting from the heaviest nucleus with $A=56$ (\Fe) 
and then proceeds downward in mass. To account for the decay $^{10}$\Be$\rightarrow\,^{10}$\B, the loop is repeated twice.
In this work I adopt a set of fragmentation cross-sections determined in \citet{Tomassetti2012Isotopes} for the production of 
$^{2}$\textsf{H} and $^{3}$\textsf{He}, and those of \citet{Tomassetti2015XS} for the production of \Li-\Be-\B nuclei.
The last term of Eq.\,\ref{Eq::PropagationEquation} describes Coulomb and ionization losses by means of the momentum loss rate $\dot{p}_{j}=dp_{j}/dt$.
Energy changes may also include diffusive reacceleration, which is described under \texttt{GALPROP} as a diffusion process in momentum space.  
While reacceleration have been successful in reproducing the peak of the \BC{} ratio,
it encounters problems for primary elements such as \p and \He in the GeV energy region \citep{Moskalenko2003}. 
Since no consensus on diffusive reacceleration has been reached for this regime, 
this mechanism will be disregarded in the following.
The diffusion coefficient $D= D(\textbf{r},p)$ is taken as spatially homogeneous and rigidity dependent:
$D(R)=\beta^{\eta} D_{0}\left(R/R_{0}\right)^{\delta}$, where $D_{0}$ fixes the normalization at the 
reference rigidity $R_{0}$, and the parameter $\delta$ specifies its rigidity dependence. 
I employ a plain diffusion model with Iroshnikov-Kraichnan spectrum with $\delta=1/2$, and 
$\eta$ is set to $-1$ to effectively account for a faster diffusion at sub-GV rigidities. 
These parameters have been cross-checked against data on the \BC{} and $^{3}$\He/$^{4}$\He ratios \citep{Aguilar2011}, from sub-GeV to TeV energies.
The transport equation is solved for steady-state condition $\partial \mathcal{N}_{j} / \partial t = 0$ into
a cylindrical diffusion region of radius $r_{\rm max}=$30\,kpc and half-thickness $L=$\,5\,kpc, 
with boundary conditions $\mathcal{N}_{j}$($r$=$r_{\rm max}$)=0 and $\mathcal{N}_{j}$($z$=$\pm$$L$)=0. 
The local IS flux is therefore computed for each particle species at the coordinates ${r_{\odot}}=8.3\,kpc$ and $z=0$:
\begin{equation}\label{Eq::ISFlux}
  \psi^{\rm IS}_{j}(E) = \frac{c A}{4\pi}\mathcal{N}_{j}(r_{\odot},p) \,,
\end{equation}
where $A$ is the mass number, and the flux $\psi_{j}^{\rm IS}$ is eventually 
expressed as function of kinetic energy per nucleon $E$.
The model predictions from this setup are discussed Sect.\,\ref{Sec::Results}.

\section{Solar Modulation}    
\label{Sec::SolarModulation}  

Solar modulation of CRs in the heliosphere is an important effect
that limits our ability 
in understanding their acceleration and propagation processes in the Galaxy.
The CR transport in the heliosphere is determined by spatial diffusion, adiabatic energy loss,
convection with the solar wind, gradient, curvature and current sheet drift effects \citep{Potgieter2013,Potgieter2014b}. 
In addition to precision data collected in low Earth orbit, invaluable data of the Voyager missions became available 
in the last years. The Voyager-1 spacecraft appeared to have crossed the solar termination shock and the heliopause, 
and it is now sampling the interstellar space, providing us with the very first data of \emph{unmodulated} IS fluxes of CRs.
The Voyager-1 data have been recently used in a number of CR physics 
studies \citep{Aloisio2015,Bisschoff2016,Corti2016,Webber2015,WebberVilla2016,Manuel2014,Potgieter2014,Cholis2016,Ghelfi2016,Schlickeiser2014,Ptuskin2015}. 

In the description CR data from ground-based or low-Earth-orbit experiment, 
the so-called \emph{force-field} (FF) approximation is widely used in CR physics thanks to its simplicity \citep{Gleeson1968,Caballero2004}. 
It comes from a steady-state solution of the Parker's equation for a spherically symmetric solar wind
and an fully isotropic diffusion coefficient.
The FF method provides an analytical one-to-one correspondence between arrival (top-of-atmosphere) 
and IS fluxes in terms of a lower shift in energy and flux of the IS quantities. 
For a given $Z$-charged CR species at given epoch in the course of the 11-year cycle, 
the energy is shifted via the relation $E^{\odot}=E^{\rm IS} - \frac{|Z|}{A} \phi$, while the modulated flux is given by:
\begin{equation}\label{Eq::ForceField}
\psi^{\odot} = \frac{(E+ mc^{2})^{2}- m^{2}c^{4}}{(E+mc^{2} +|Z|e\phi)^{2}-m^{2}c^{4}} \times \psi^{\rm IS}(E + |Z|e\phi)
\end{equation}
The solar activity conditions are expressed in terms of only the modulation potential $\phi$, which has the dimension of 
an electric potential or a rigidity. The parameter $\phi$ is generally interpreted as the average energy loss (per charge unit)
experienced by charged particles traversing the heliosphere. The long-term temporal variation of the solar modulation effect is
therefore expressed in term of a time-dependent parameter $\phi=\phi(t)$. 
Several strategies have been developed for the reconstruction of the modulation level $\phi(t)$ at different epochs \citep{Ghelfi2016,Usoskin2011}. 
Important limitations of the FF model are its lack of predictive power and the fact that FF-based results depend on the assumed IS flux. 
However, in spite of a large proliferation of IS models proposed in the last decades \citep{Usoskin2005,Usoskin2011},
recent Voyager 1 data can now provide useful constraints to the low-energy shape of IS spectra \citet{Stone2013}.
Clearly, the FF approach neglects fundamental transport processes such as drift motion, adiabatic cooling, 
or the tensor nature of the particle diffusion, that are accounted in more advanced formulations
\citep{Kappl2016,Maccione2013,Strauss2011,DellaTorre2012,Potgieter2014Pamela,Potgieter2014b}.

\subsection{Solar modulation for AMS in space} 

The long exposure of new-generation experiments such as \AMS or PAMELA gives rise to another concern.
These measurements are provided over very long observation times $\Delta T$
during which the solar activity evolves appreciably. 
Equation\,\ref{Eq::ForceField} can be re-written as a transformation of the type:
\begin{equation}\label{Eq::Operator}
\psi^{\odot} = \mathcal{\hat{G}}_{t}\left[\psi^{\rm IS}\right] 
\end{equation}
Following the above equation, the application of the FF method to the \AMS data would require the derivation of
an effective value for the parameter $\langle\phi\rangle$, averaged over an observation time of $\Delta T\approx$\,2.5\,yrs.
During this observation time, the number of $j$--type particles detected by \AMS at energy $E$ between $E_{1}$ and $E_{2}$ can estimated as \citep{Tomassetti2015XS}:
\begin{equation}\label{Eq::DetectorModel}
\Delta N_{j} = \int_{E_{1}}^{E_{2}} dE \int_{T_{1}}^{T_{2}} \psi^{\odot}_{j}(t,E) \cdot \mathcal{A}_{j}(E) \cdot \mathcal{H}_{j}(t,E)\cdot dt \,,
\end{equation}
which describes the convolution of the arrival flux $\psi^{\odot}_{j}(t,E)$ with the total detector acceptance $\mathcal{A}_{j}$.
All relevant efficiencies are highly stable with time so that the acceptance can be taken as time independent \citep{Aguilar2015Proton}.
The function $\mathcal{H}_{j}$ is the \emph{geomagnetic transmission function} for Galactic CRs.
For all CR particles, it can be modeled as a universally rigidity-dependent smoothed step function, $\mathcal{H}= H(R-R_{C})$ (or a sigmoid function).
The particle-dependence of the $\mathcal{H}$ function arises from the $Z/A$-dependent conversion $R\rightarrow E$,
while its time-dependence is implicit in the \emph{geomagnetic cutoff} rigidity, $R_{\rm C}(t)\sim$\,0.2--30\,GV, that changes rapidly with the ISS orbit \citep{SmartShea2005}.
It should be noted that the temporal variation of $\mathcal{H}$ (arising from the 91\,min orbital period of the ISS) 
is much faster than the flux variation due to long-term solar modulation (occurring on monthly time-scales).
Hence I can consider the factorized integral $\mathcal{T}_{j}(E) \equiv \int_{\Delta T} \alpha(t) \mathcal{H}_{j}(t,E) dt$,
where the factor $\alpha\approx$\,95\,\% is included to account for the detector live-time \citep{Tomassetti2015XS}.
Furthermore, one can simplify $\int_{E1}^{E2} [\dots] dE \cong [\dots]\Delta E$ for all considered energy bins,
so that the total energy-binned flux measured by \AMS can be expressed as:
\begin{equation}\label{Eq::MeasuredFlux}
\langle \psi_{j}^{\odot}(E) \rangle \cong \frac{\Delta N_{j}}{\mathcal{T}_{j}(E) \mathcal{A}_{j}(E) \Delta E} \approx \frac{1}{\Delta T}\int_{T_{1}}^{T_{2}}\psi^{\odot}_{j}(t,E) dt
\end{equation}
This equation shows that the measured flux can be interpreted as a time-averaged quantity,
$\langle \psi^{\odot} \rangle \equiv \frac{1}{T}\sum_{k}\psi^{\odot}(t_{k},E)\delta t_{k}$, 
where $\psi^{\odot}(t_{k})$ is the flux at $t=t_{k}$ evaluated in the time interval $\delta t_{k}$. 
As discussed, the interval $\delta t_{k}$ has to be larger than the ISS orbital period $T_{\rm ISS}=91$\,min 
and small enough to appreciate the flux variation of $\psi^{\odot}$ in connection the long-term modulation. 
A reasonable choice is $\delta t_{k}\approx$\,1\,month.
To model the CR solar modulation, I will adopt the following approach:
\begin{equation}\label{Eq::NonLinearity}
\psi^{\odot} = \frac{1}{T}\sum_{k} \delta t_{k} \mathcal{\hat{G}}_{t_{k}}\left[ \psi^{\rm IS}\right] 
\end{equation}
It is important to realize that Eq.\,\ref{Eq::Operator} and Eq.\,\ref{Eq::NonLinearity} are not mathematically equivalent, 
reflecting the fact that $\mathcal{\hat{G}}$ is a nonlinear operator. 
Hence the usual application of the FF method as in Eq.\,\ref{Eq::Operator} may not give correct results.

\subsection{Determination of the modulation potential} 

The use of an approach based on Eq.\,\ref{Eq::NonLinearity} requires a reconstruction of the
time evolution of the modulation potential $\phi=\phi(t)$ over the period of interest. I proceed as follows: 
\begin{enumerate}
\item 
  With the use of NM counting rates,
  I perform a reconstruction of the FF modulation potential $\phi$ and its evolution 
  at different epochs, on a monthly basis, characterized by different 
  strength of the solar modulation effect.
\item 
  Once the NM-driven reconstruction of the parameter $\phi$ is established, 
  I apply Eq.\,\ref{Eq::NonLinearity} to the IS fluxes calculated in Sect.\,\ref{Sec::CRPropagation}, 
  in order obtain the solar-modulated flux for the cumulated period of \AMS observations.
\end{enumerate}
%
\begin{figure*}[!t]
\begin{center}
\includegraphics[width=2\columnwidth]{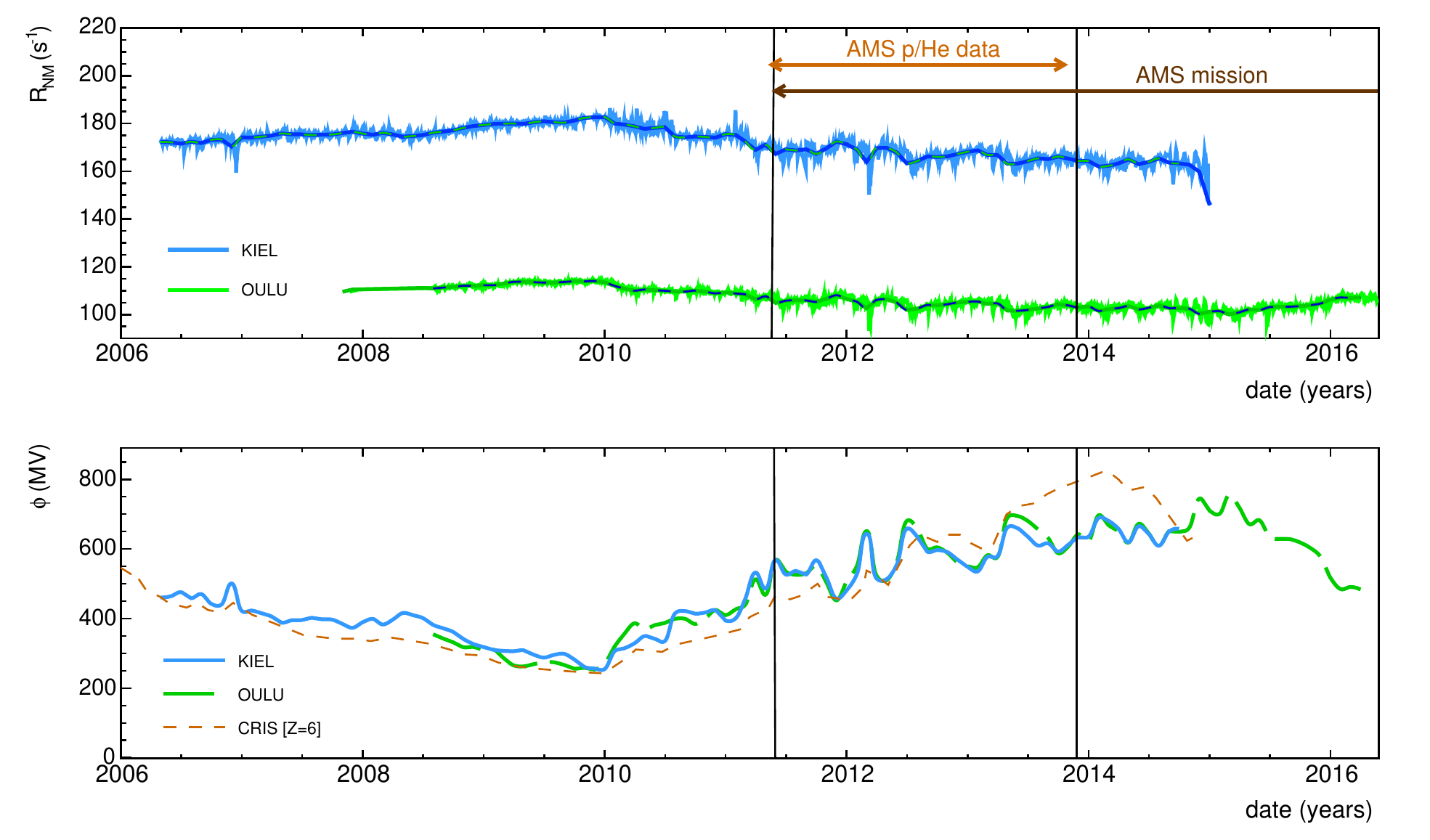}
\end{center}
\caption{
Top: NM counting rates as function of time from the Oulu and Kiel stations.
The rates are shown in term of daily (thin lines) and monthly-averaged counts (thick lines) 
corrected for detector efficiency and pressure.
The calculated rates are superimposed (thin dashed lines).
Bottom: time evolution of the solar modulation potential $\phi$ as reconstructed using Oulu and Kiev NM data in term of monthly average rates. 
The $\phi$ reconstruction operated by the CRIS/ACE Collaboration using the CR carbon flux is also shown (thin dashed line).
}
\label{Fig::ccNM}
\end{figure*}
%
NM detectors consist in a worldwide network of ground-based particle counters.
These detectors measure the secondary particles of hadronic showers generated by CR interactions
with the atmosphere \citep{Dorman2009}.
The counters are in general surrounded by thick layers of lead, in order to enhance the
detected signal by the production of neutrons generated by nucleons or muons traversing these layers.
For a ground-based NM detector $d$, the link between the count rate $\mathcal{R}_{\rm NM}^{d}$ and the CR fluxes at the top of 
the atmosphere $\psi_{j}^{\odot}$ (with $j=$\p,\,\He) can be expressed by:
\begin{equation}\label{Eq::NMRate}
\mathcal{R}_{\rm NM}^{d}(t) = \int_{0}^{\infty}  dE \cdot \sum_{j={\rm CRs}} \mathcal{H}^{d}_{j}(E)\cdot\mathcal{Y}^{d}_{j}(t,E)\cdot\psi^{\odot}_{j}(t,E) 
\end{equation}
The transmission function $\mathcal{H}^{d}$ is parameterized as a smoothed Heaviside function of rigidity \citep{Tomassetti2015XS}
and assumed to be time-independent, \ie, for a given detector location, the rigidity cutoff $R_{C}$ is assumed to be constant.
The quantity $\mathcal{Y}^{d}_{j}$ represents the response function, in units of m$^{2}$\,sr, for a $j$-type primary CR species at energy $E$. 
In this work, I simply express the yield function as:
\begin{equation}\label{Eq::Vortex}
\mathcal{Y}^{d}_{j} = \mathcal{V}^{d}\, .\, \mathcal{F}^{d}_{j} \,,
\end{equation}
where $\mathcal{F}^{d}_{j}(t,E)$ accounts for all time/energy dependencies, including hadron physics models for secondary production,
while the factor $\mathcal{V}^{d} \propto exp(f_{d}h_{d})$ expresses the absolute normalization of the detector response and its dependence on the altitude $h_{d}$,
To model $\mathcal{Y}^{d}_{j}$, the parameterization proposed in \citet{Maurin2015} is adopted.
It describes nucleon/muon production generated by CR protons and helium showers \citep{Cheminet2013}. The small contribution from heavier CR nuclei is disregarded.
%
\begin{figure*}[!t]
\begin{center}
\includegraphics[width=\columnwidth]{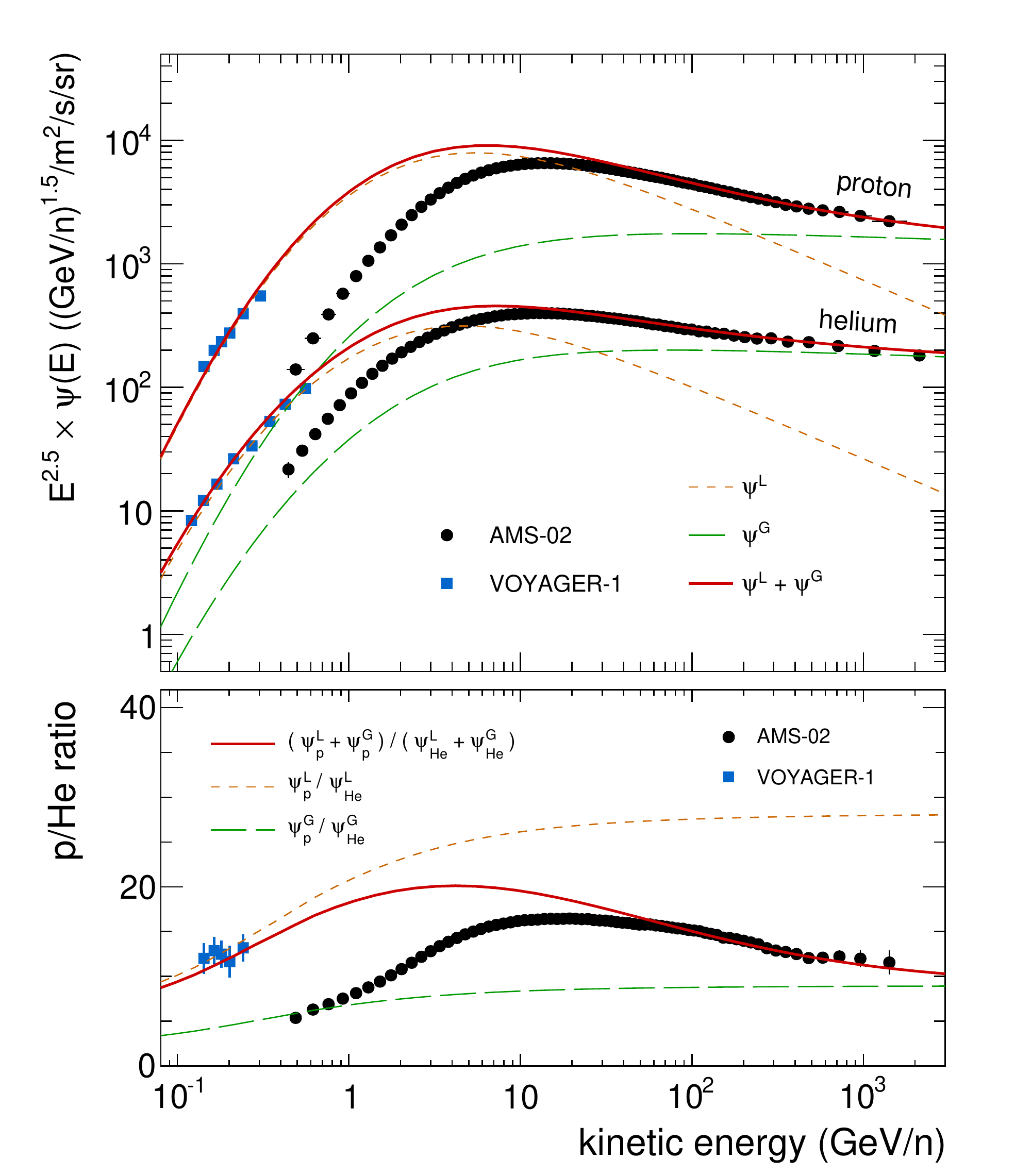}
\includegraphics[width=\columnwidth]{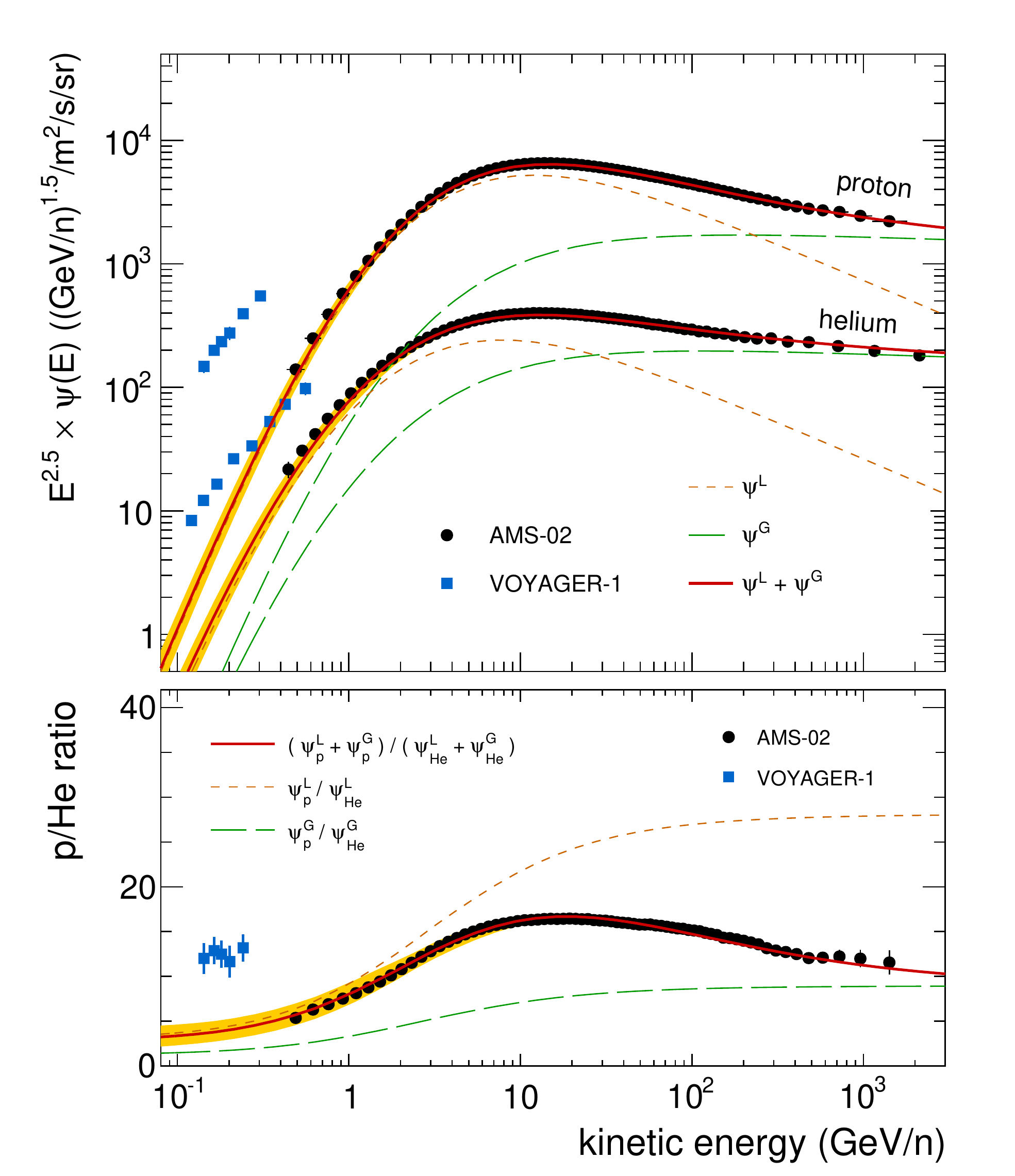}
\end{center}
\caption{
Top: the proton and helium fluxes as function of kinetic energy per nucleon. The model predictions (solid lines) are shown
in comparisons with new data from \AMS \citep{Aguilar2015Proton,Aguilar2015Helium} and from Voyager-1 \citep{Stone2013}. 
Bottom: the \pHe ratio as function of kinetic energy per nucleon. The \AMS data of \p and \He have
been converted from rigidity and then interpolated to a common energy grid, as in \citet{Tomassetti2015PHeRatio}.
The dashed lines reflect the flux ratio for the single source components. The left panels the model predictions
are shown for the IS fluxes. In the right, the fluxes and the \pHe ratio have been modulated with the procedure 
described in Sect.\,\ref{Sec::SolarModulation}. The yellow bands are the uncertainties associated with solar the modulation modeling.
}
\label{Fig::ccProtonHeliumMOD}
\end{figure*}
%
I consider the measured monthly-averaged rates from NM stations
in Oulu ($h=$\,15\,m, $R_{C}=$\,0.81\,GV) and Kiel ($h=$\,54\,m, $R_{C}=$\,2.3\,GV) \citep{Steigies2015}.
For each of the 120 months between July 2006 and January 2016, the convolution of Eq.\,\ref{Eq::NMRate} is calculated 
using my IS flux predictions as input and the modulation potential $\phi$ as free parameter.
The parameter  $\phi$ is obtained from the request of agreement between
the \emph{calculated} rate $\mathcal{\tilde{R}}^{d}$ and the \emph{measured} rate $\mathcal{R}^{d}$,
together with the requirement that $\int_{\Delta T^{d}}\tilde{R}^{d}(t) dt = \int_{\Delta T^{d}} R^{d}(t) dt$
where the integrals run over the considered observation periods $\Delta T^{d}$.
In practice, for a given NM station, the determination of $\phi(t)$ relies on an iterative procedure
consisting of an adaptive grid scan on the parameter $\phi$, for each time interval, until the quantity
\begin{equation}
d^{d} = \left| \frac{\mathcal{\tilde{R}}^{d}(\phi) - \mathcal{{R}}^{d}}{\mathcal{{R}}^{d}} \right|
\end{equation}
gets its minimum value or falls below the value of $10^{-3}$, for which I declare that the convergence is reached.
The results of this procedure are shown in Fig.\,\ref{Fig::ccNM} for both Oulu and Kiel stations. 
The top panel shows the counting rates $\mathcal{R}$ as function of time.
The data, corrected for detector efficiency and pressure, are shown in term of daily and monthly-averaged counts (solid lines).
The calculated rates $\mathcal{\tilde{R}}$ are superimposed in the figure (thin dashed lines), but they agree with $\mathcal{R}$ 
at the level $\mathcal{O}(10^{-3})$. 

The bottom panel shows the resulting modulation parameter $\phi=\phi(t)$, determined by each fit on monthly basis, 
using the data from both detectors. The time is expressed in units of fractional years.
As expected, the inferred modulation potential shows an anti-correlation relationship with the NM ratios.
The difference between the two reconstructions reflects systematic uncertainties in the knowledge of the two detector responses. 
They are associated to an uncertainty on $\phi$ of $\sim$\,25\,MV.
In the figure it is also shown the $\phi$ reconstruction determined by the CRIS/ACE Collaboration, obtained using carbon data, 
updated to January 2015 \citep[see][]{Wiedenbeck2009}.
While the agreement on the overall trend is fairly good, large discrepancies appears during times of high solar activity. 
It should be noted that the CRIS-driven reconstruction has not been made under the FF approximation.
The use of different input spectra, arising from different CR propagation models, is another possible reason of this discrepancy:
contrary to proton and \He, the IS flux of carbon is poorly known.
I also stress that the CRIS analysis is performed on data at $E\sim$\,50-500\,MeV/nucleon.
Data at this energy are very useful for solar modulation studies, but they significantly differ 
from the average CR energy probed by NMs ($\sim$\,2-20\,GeV/n),
because the latter consists in energy-integrated rates of CRs at rigidities greater than $R_{C}$. 
Discrepancies in these two energy regions were previously noticed \citep{Gieseler2015,Wiedenbeck2009}.
It is also interesting to note that the CRIS data are not subjected to magnetospheric effects, while the NR response 
may suffer from unaccounted variations of the local geomagnetic field (unaccounted because $R^{d}_{C}$ is assumed to be constant). 
The \AMS data collection time for \p and \He is indicated in Fig.\,\ref{Fig::ccNM}. The \AMS mission is characterized by a high level 
of solar activity in comparison to the pre-\AMS region where a long solar minimum occurred.
Equation\,\ref{Eq::NonLinearity} is used to compute the modulated fluxes $\psi^{\odot}$ of CR protons and helium
predicted from my model over the observation period. This period is covered by both NM detectors, hence 
I make use of the average value of the two corresponding parameters $\phi$.
For each $k$-th month and each $j$-th species, the uncertainty on the predicted modulation potential $\delta\phi$ is translated into 
an uncertainty on the corresponding modulated flux:
\begin{equation}
 \delta\psi_{j}^{\odot}(t_{k})=\delta\mathcal{\hat{G}}_{t_{k}}[\psi_{j}^{\rm IS}].
\end{equation}
The final uncertainty on the total modulated flux is computed, from Eq.\,\ref{Eq::NonLinearity}, by assuming maximum error correlation.

\section{Results and discussion}  
\label{Sec::Results}              

\subsection{Testing the two-component model}  

The main results are illustrated in Fig.\,\ref{Fig::ccProtonHeliumMOD}. 
The model calculations of proton and \He are plotted as function of kinetic energy per nucleon in comparison with the data. 
The proton and the helium fluxes are shown (top), together with their ratios (bottom), 
before (left) and after (right) the application of solar modulation as described in Sect.\,\ref{Sec::SolarModulation}. 
The \AMS data of \p and \He have been converted from rigidity into kinetic energy per nucleon.
Thus the \pHe ratio data have been calculated after a log-linear interpolation into a common energy grid.
The energy spectra of both species are multiplied by $E^{2.5}$.
The fluxes arising from the two source components $L$ and $G$ are also shown as dashed lines.
The IS fluxes shown in the left panel are tuned to match the Voyager-1 data at low energy and the \AMS data at energy above $\sim$\,10\,GeV/nucleon. 
The flux transition between the two components produces a progressive change in slope of the spectra at $E\gtrsim$\,100\,GeV/n and 
a smooth decrease of the \pHe ratio (bottom panel) in the region $E\sim$\,10\,--\,1000\,GeV/n, in good accordance with the data.
In this model, the predicted \pHe{} ratio can be written as the ratio 
$\psi_{\rm p}/\psi_{\rm He}\equiv (\psi_{\rm p}^{\rm L} + \psi_{\rm p}^{\rm G})/(\psi_{\rm He}^{\rm L} + \psi_{\rm He}^{\rm G})$, 
where the model ratios corresponding to the single source components are
$\psi_{\rm p}^{\rm L}/\psi_{\rm He}^{\rm L}$ (short-dashed lines), and 
$\psi_{\rm p}^{\rm G}/\psi_{\rm He}^{\rm G}$ (long-dashed lines). 
In the sub-GeV energy region, the TCM predictions reflect the properties of the former component. 
The figure shows a good agreement between calculations and Voyager-1 data, supporting the hypothesis that the primary CR 
flux observed at these energies is generated by a unique type of accelerator with elemental-independent injection spectrum.
From the model, in fact, one has $\psi_{\rm p}\approx \psi_{\rm p}^{\rm L}$ and  $\psi_{\rm He}\approx \psi_{\rm He}^{\rm L}$.
In other related works, it was suggested that the CR flux in this energy region may be dominated by nearby SNR explosions 
possibly occurred in relatively recent epochs and associated with the local 
bubble \citep{TomassettiDonato2015,Tomassetti2015Upturn,ErlykinWolfendale2012,Malkov2012,Moskalenko2003}.
This idea is also supported by recent numerical simulations beyond the usual steady-state approximation \citep{Kachelriess2015}.
Within the TCM, one can also argue that local component is characterized by relatively steep injection spectra and low metallicity.

In the right panel of Fig.\,\ref{Fig::ccProtonHeliumMOD}, the fluxes have been solar modulated over the \AMS observation period.
The yellow bands describe the uncertainties associated with the modulation modeling.
The model comparison with the \AMS data on the \pHe ratio is very good. It is important to stress that
several TCM parameters are not determined from direct fits to the \AMS data:
the effect of modulation modeling is based on NM counting rates, the relative abundance of \Hyd{} and \He{} is fixed by Voyager-1 data,
and the injection indices of the low-energy sources are imposed to be the same for all primary particles.
Thus, the parameters
$q_{\rm He}^{0}/q_{\rm H}^{0}$, $\nu_{\rm He}/\nu_{\rm p}$, and the time-series $\phi_{k}=\phi(t_{k})$ are not forced to agree with the \AMS data.
Nonetheless, the comparison gives $\chi^{2}/22\approx$\,1.35 ($\chi^{2}/25\approx$\,1.15) for the proton (helium) data points below 10 GeV/n.

It should also be noted that several studies claimed good levels of agreement, on both \p and \He{} data,
using standard injection models based on one universal class of CR sources.
In these models the low-energy flux can be determined by various physics processes.
For example, \citet{Aloisio2015} proposed a scenario where the TeV spectral hardening is caused by a transition 
between diffusion in CR self-induced turbulence and diffusion in pre-existing turbulence, while
in the low-energy flux is strongly affected by CR advection with the self-generated waves. 
In \citet{Bisschoff2016} it was shown that the incorporation of diffusive reacceleration 
may allow for a consistent description of proton, \He, and \C interstellar spectra as measured by Voyager-1. 
In \citet{Tomassetti2015TwoHalo} I have proposed a ``two-halo'' model of CR propagation where
high-energy spectra at the TeV scale are dominated by CR diffusion in proximity of the Galactic plane (within $d\lesssim$\,600\,pc), 
while the low-energy shape of the interstellar flux is determined by the CR diffusion properties in the outer halo (up to $L\sim$\,5\,kpc).
A common feature of the above scenarios is that the injection spectral indices $\nu$ of proton and helium
must be different by a factor $\sim$\,0.07-0.1 to match the observations. This demands some
fundamental revisitation of in diffusive shock acceleration mechanisms \citep{Serpico2015}
which is not necessary in the TCM scenario presented here.

\subsection{On the solar modulation modeling} 

\begin{figure}[!t]
\begin{center}
\includegraphics[width=\columnwidth]{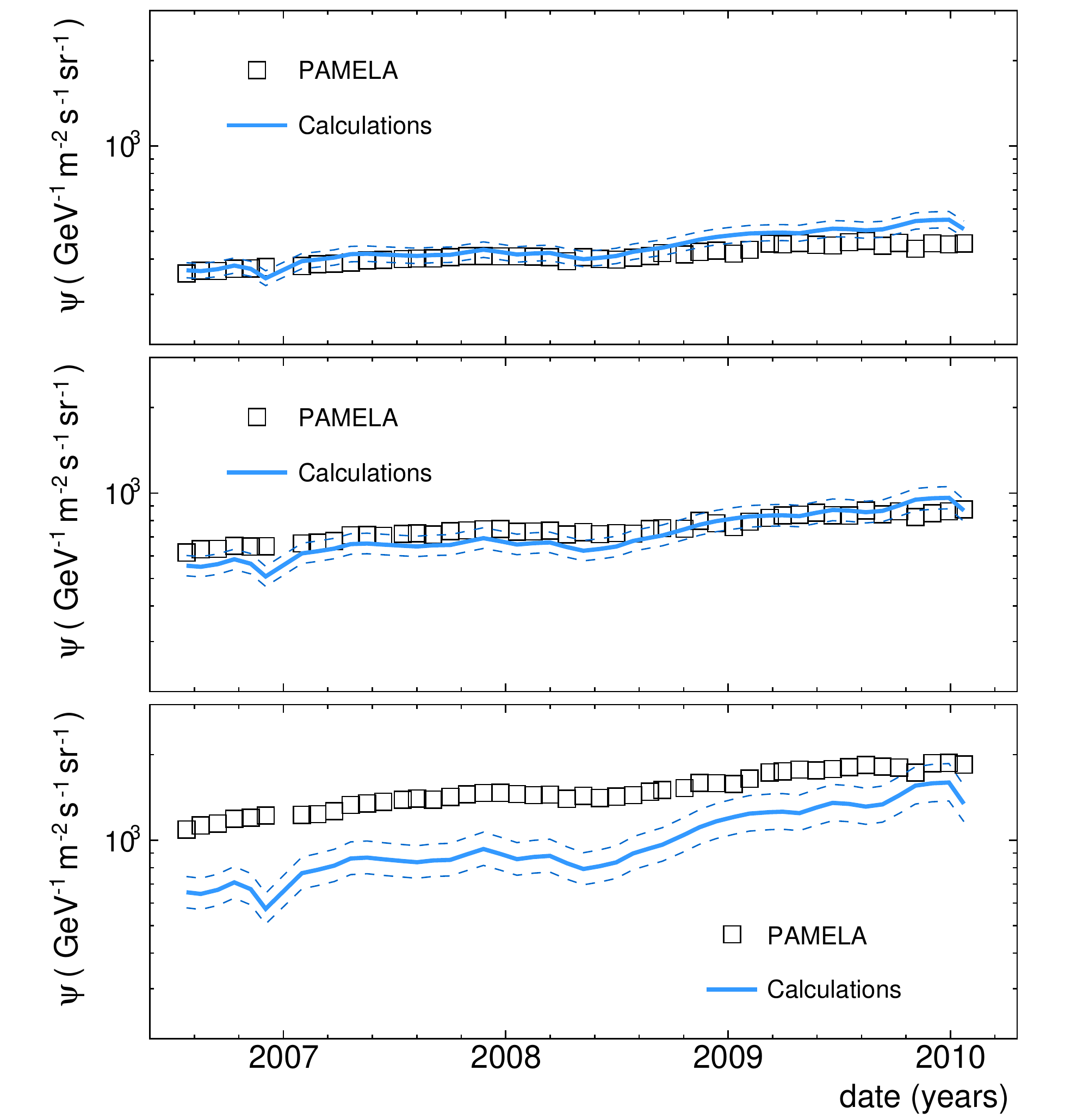}
\end{center}
\caption{
Monthly proton fluxes measured from the PAMELA experiment between 2006 and 2010 at 
different energies (top to bottom: $\sim$\,2\,GeV, $\sim$\,1\,GeV, and $\sim$\,0.4\,GeV) in comparisons
with TCM calculations, where the modulation was calibrated using NM data from the Kiel station.
The dashed lines indicate the uncertainties in the calculations.
}\label{Fig::ccPamelaProton}
\end{figure}

\begin{figure}[!t]
\begin{center}
\includegraphics[width=0.9\columnwidth]{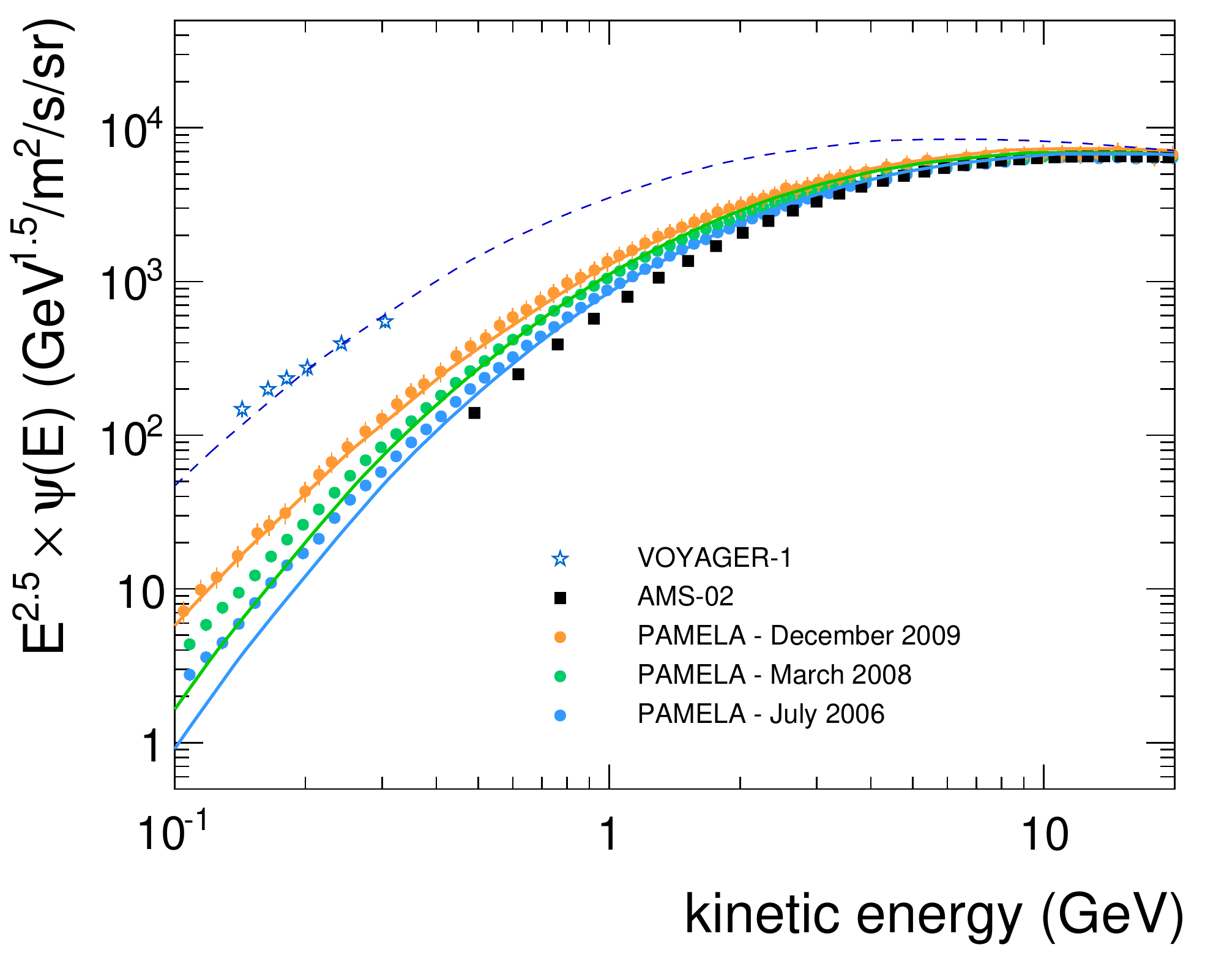}
\end{center}
\caption{
Differential proton fluxes measured from the PAMELA experiment on December 2006, March 2008, and July 2010
compared with TCM calculations. The modulation parameter is calibrated using NM data from Kiel. 
Data from \AMS and Voyager 1 are also shown, along with the interstellar TCM flux.
}\label{Fig::ccPamelaVSAMSModulation}
\end{figure}

Another point of discussion the discrepancy noted in Sect.\,\ref{Sec::SolarModulation}
between the CRIS-driven and the NM-driven modulation parameter reconstruction \citep[see also][]{Gieseler2015}.
In the top panel of Fig.\,\ref{Fig::ccProtonHeliumMOD} it can be seen that 
both \p and \He fluxes are slightly under-estimated at kinetic energy of about 0.5\,GeV/n.
This discrepancy can be inspected using low-energy proton data
measured by the PAMELA experiment from July 2006 to January 2010 \citep{Adriani2013,Potgieter2014Pamela}.
Figure\,\ref{Fig::ccPamelaProton} shows the comparison between proton flux calculations and
data from PAMELA at three kinetic energy windows ($\sim$\,1.8-2.2, 0.8-1.2, and 0.2-0.6\,GeV, from
top to bottom). 
In Fig.\,\ref{Fig::ccPamelaVSAMSModulation}, TCM calculations are compared with the PAMELA fluxes observed in three distinct epochs,
along with the data from Voyager 1 and \AMS.
In the lowest energy region, it can be seen that the fluxes are systematically underpredicted by the model,
and this discrepancy increases with increasing level of solar modulation. 
In terms of the FF parameter $\phi$, the level of disagreement is approximately $\sim$\,50\,MeV at $E\lesssim$\,0.5\,GeV.
Hence the observed difference between NM-driven and CRIS-driven modulation parameters (Fig.\,\ref{Fig::ccNM}) 
is probably ascribed to solar modulation modeling rather than to the choice of IS spectra.
%
%
%
To make a consistent description of CR data down to $E\sim$\,50\,MeV/n, 
a more refined solar modulation modeling is required, certainly beyond the FF approximation.


\subsection{On the CR propagation parameters} 

The determination of acceleration and transport parameters is somewhat tricky in CR propagation
and deserves to be discussed. In principle, linear and steady-state calculations of 
diffusive shock acceleration predict universal spectra with $\nu\equiv\,2$ for strong shocks. 
However, instantaneous spectra and SNR shock properties are expected to evolve with time.
It follows that a convolution of particle acceleration (and escape) over the SNR evolution
gives a total spectrum which may be steeper than $E^{-2}$.
This realization is in accordance with recent observations of $\gamma$-ray spectra from several SNRs 
such as Tycho, W44, or IC-443 \citep{Caprioli2012,Tomassetti2015Upturn}.
Thus, in CR propagation models with steady-state source terms ($q^{\rm pri} \propto R^{-\nu}$), the parameter $\nu$
is regarded as an effective quantity, to be determined from the data, representing the average SNR properties
(and/or their average acceleration histories).
%
\begin{figure}[!t]
\begin{center}
\includegraphics[width=\columnwidth]{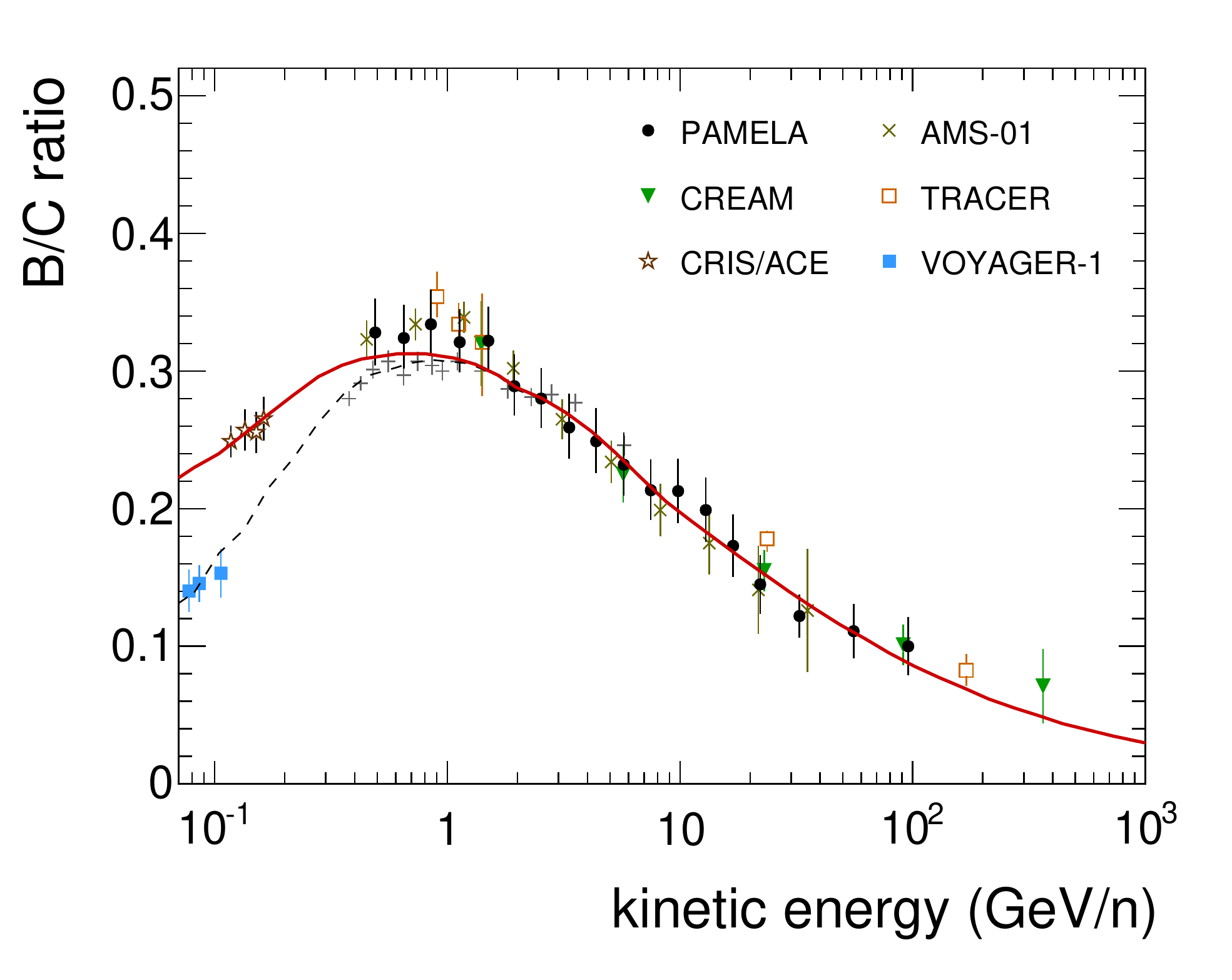}
\end{center}
\caption{
  TCM calculations for the interstellar (solid) and modulated (dashed lines) \BC{} ratio
  in comparison with existing measurements \citep{Adriani2014,Aguilar2010,Ahn2008,DeNolfo2006,Obermeier2011,WebberVilla2016}.
  The solar modulation level is simply set at the level $\phi=300$\,MV. The diffusion coefficient is taken of the form
  $D\propto \beta^{\eta} R^{\delta}$ with scaling indices $\delta=1/2$ and $\eta=-1$. Fragmentation cross-sections for 
  boron production are taken from \citet{Tomassetti2015XS}.
}\label{Fig::ccBCRatio}
\end{figure}
%
Besides, its experimental determination
is affected by strong degeneracies between acceleration and transport parameters.
Since the spectra of primary nuclei are only sensitive to $\gamma\approx\nu+\delta$,
knowledge of the parameter $\delta$ is essential.
Theoretically favored values are $\delta=1/3$ and $\delta=1/2$, which correspond to a
Kolmogorov and Iroshnikov-Kraichnan type spectrum of interstellar turbulence, respectively.
Data on the \BC{} ratio are widely used to constraints the parameter $\delta$.
The \BC{} ratio from the TCM is plotted in Fig.\,\ref{Fig::ccBCRatio} in comparison with the recent measurements.
As shown, the choice of $\delta=1/2$ is well consistent with the new data from PAMELA.
Under this setting no reacceleration is needed. The 1 GeV/n peak of the \BC{} ratio is interpreted as a
low-rigidity modification of the diffusion coefficient (as expected \eg{} from wave damping, see \citet{Ptuskin2006}) 
which is described by the parameter $\eta$.
This scenario also is consistent with antiproton data \citep{DiBernardo2010}.
In contrast, reacceleration models with Kolmogorov diffusion and Alfv\'enic speed $v_{A}\sim\,30-40$\,km\,s$^{-1}$
reproduce very well the sharp peak in the \BC{} ratio, but they require the introduction of
artificial breaks in injection at $E\lesssim$\,10\,GeV, in order to avoid the development
of unphysical bumps. Furthermore, these models are known to underpredict the antiproton flux 
by $\sim$\,30\,\% at $\sim$\,1-10\,GeV of energy \citep{Moskalenko2002,Moskalenko2003,Grenier2015}.
Given the important role of CR antiprotons in dark matter searches, the discrimination among the two models is crucial.
With the currently available data it is not possible to achieve such a discrimination.
Fortunately, we are in a proficient era for the CR physics investigation. 
Accurate measurements of the \BC{} ratio at the TeV energy scale are expected soon from several ongoing space missions \citep{Maestro2015}.

\section{Conclusions}      
\label{Sec::Conclusions}   

This work is motivated by the search of a comprehensive model for Galactic CRs 
that is able to account for the many puzzling anomalies recently discovered in their energy spectrum.
Some of the observed features seem to suggest that the total CR flux observed at Earth is provided by different types of accelerators. 
As shown in earlier works, a TCM scenario is able to accounts for several important features in the CR spectrum
such as the rise of the positron fraction \citep{TomassettiDonato2015} or the observed upturn in light/heavy nuclear ratios \citep{Tomassetti2015Upturn}.
Regarding the unexplained decrease of the \pHe{} ratio, the TCM hypothesis of having two types of contributing SNRs
($L$ and $G$) is essential to reconcile the observations with the universality of CR acceleration 
(\ie, the circumstance that CR sources provide composition-blind injection spectra).
This paper is mainly focused on the low-energy part of the spectrum of CR protons and helium nuclei ($E\sim$0.1--10\,GeV)
where the contribution of the $G$-component is expected to become negligible.
At these energies, however, making predictions for the CR flux at Earth requires to account for the solar modulation effect.
Within the FF approximation, I have determined the time-dependence of the modulation parameter using complementary sets of data provided by NMs. 
I have addressed the problem of effectively modeling solar modulation for describing CR data collected over large observation periods. 
I have used a simple procedure based on Eq.\,\ref{Eq::NonLinearity} 
in order to account for the temporal evolution of the modulation potential over the period of data observation.
This method has general validity and should be adopted for describing the data reported by long-exposure CR 
detection experiments, provided that their collection power is stable with time.
As shown, the recent \p-\He data reported by \AMS and Voyager-1 are in good agreement with the TCM predictions.  
On the other hand, discrepancies were found in the energy region $E\lesssim$\,0.5\,GeV/n that are 
presumably due to a breakdown of the solar modulation modeling based on the FF approximation.
Further improvements on this side require 
to incorporate other known processes that rule the solar modulation of CRs in the heliosphere.

\section*{Acknowledgements} 
%
I would like to thank my colleagues of the \AMS time dependence study group
and in particular the authors of \citet{Ghelfi2016} and \citet{Corti2016} 
for discussions. I thank Veronica Bindi for the invitation at the
workshop on \emph{Solar Energetic Particles, Solar Modulation and Space Radiation} in Honolulu,
and the anonimous reviewers for their helpful comments. 
Support from MAtISSE is acknowledged.
This project has received funding from the European Union's Horizon 2020 research and innovation programme under the Marie Sklodowska-Curie grant agreement No 707543.
\\



\end{document}